\documentclass[twocolumn,showpacs,preprintnumbers,amsmath,amssymb]{revtex4}

\usepackage{graphicx}
\usepackage{dcolumn}
\usepackage{bm}
\usepackage{color}

\begin{document}


\title{Conductance asymmetry of graphene pn junction }

\author{Tony Low, Seokmin Hong, Joerg Appenzeller, Supriyo Datta and Mark Lundstrom}
\affiliation{Network for Computational Nanotechnology\\
Birck Nanotechnology Center\\
School of Electrical \& Computer Engineering, Purdue University, West Lafayette, IN47906, USA }%
\email{tonyaslow@gmail.com}

\date{\today}

\begin{abstract}

We use the non-equilibrium Green function (NEGF) method in the ballistic limit to provide a quantitative description of the conductance of graphene pn junctions - an important building block for graphene electronics devices. In this paper, recent experiments on graphene junctions are explained by a ballistic transport model, but only if the finite junction transition width, $D_w$, is accounted for. In particular, the experimentally observed anamolous increase in the resistance asymmetry between nn and np junctions under low source/drain charge density conditions is also quantitatively captured by our model. In light of the requirement for sharp junctions in applications such as electron focusing, we also examine the pn junction conductance in the regime where $D_w$ is small and find that wavefunction mismatch (so-called  pseudo-spin) plays a major role in sharp pn junctions.

\end{abstract}

\maketitle
\section{\label{sec:level1}INTRODUCTION}
Graphene is a two-dimensional sheet of carbon atoms arranged in a honeycomb lattice with unique electronic properties due to its linear energy dispersion, with zero bandgap, and a spinor-like two-component wavefunction \cite{novoselov04,zhang05,semenoff84}. These characteristics give rise to interesting transport phenomena such as the absence of backscattering \cite{ando98}, anomalies in the quantum Hall regime \cite{novoselov05,zhang05}, weak anti-localization \cite{suzuura02,wu07}, so-called Klein tunneling \cite{katsnelson06}, and electron focusing analogous to optical effects that occur in negative refractive index materials \cite{cheianov07}. As such, one expects that grahene pn junctions should differ from traditional semiconductor pn junctions. The pn junction is a basic building block for electronic devices.  Developing a quantitative understanding of graphene pn junctions is an important step on the way to realizing novel devices such as graphene lenses \cite{cheianov07,shytov07} and filters \cite{katsnelson06}.  Our goal in this paper is to demonstrate that the non-equilibrium Green's function (NEGF) approach \cite{datta97} quantitatively explains recent experiments on graphene pn junctions, including the critical role of the junction depletion width and the increased in the odd resistance observed under low source/drain charge density conditions \cite{huard07}.

Electron transmission across a graphene pn junction occurs by interband tunneling. A theoretical treatment for an abrupt, graphene pn junction predicts that for a symmetric pn junction (i.e. one in which the hole and electron densities on each side of the junction are equal) the transmission probability is given by $cos^{2}\theta$, where $\theta$ is the angle between the electron's wave-vector and the normal to the junction interface \cite{katsnelson06}. Realistic pn junctions will have a transition region of finite width. For a smooth junction transition region of width $D_{w}$, the Wentzel Kramers Brillouin (WKB) approximation can be used to show that the transmission probability for a symmetric pn junction is $e^{-\pi k_{f}D_{w}sin^{2}(\theta)/2}$ \cite{cheianov06}, where $k_{f}$ is the Fermi wave-vector. Whether the transition region is abrupt or graded, the transmission is perfect when $\theta$=$0$ (i.e. commonly refered to as Klein tunneling), but the transmission decreases as $\theta$ and $D_{w}$ increase. This angular selectivity for electron transport across the pn  junction serves as a filter, allowing states with  $\left|\theta\right|$$\leq$$\sigma_{\theta}$ (where $\sigma_{\theta}$ is the spread of the angular distribution) to pass through more effectively.  The quantity $2\sigma_{\theta}$ can be viewed as the bandwidth of this filter and is what gives rise to the larger resistance of a pn junction as compared to a uniform graphene sheet.

Several research groups have recently fabricated graphene pnp devices by using electrostatic gates to create p and n regions \cite{huard07,williams07,ozyilmaz07,gorbachev08}. The typical setup consists of a back-gate and top-gate, which are used to control the amount of charge density in the source/drain and channel regions respectively.  For example, the back gates can be set to produce n-type source and drain regions and the top gate can be biased to change the middle (channel) region from n to p type.  An asymmetry in the device's source to drain resistance as a function of top-gate voltage has been experimentally observed \cite{huard07}. The amount of this resistance asymmetry is a measure of the intrinsic property of the pn junction, provided that the mean free path of carriers is larger than the transition length of the junction. One can theoretically compute the junction's transition length accounting for non-linear screening effects \cite{zhang08}.  For recent experiments \cite{huard07}, typical transition lengths for pn junction are less than $100 nm$. Recent experiments indicate that the carrier's mean free path is about $100nm$ under low temperature and moderate carrier density conditions of $10^{12}cm^{-2}$ \cite{du08}. In addition, there is experimental evidence of Fabry-Perot interference effects within the channel region in devices with channel lengths less than $100nm$ \cite{stander08,young08}, evidently a signature of phase coherent transport. Therefore, a ballistic transport model is sufficient for the study of the experimental pn junction devices reported in \cite{huard07}. 

In this paper, we present a quantitative study of the near-equilibrium IV characteristics of graphene pn junctions.  In particular, we present a systematic study of the impact of the junction transition width, $D_{w}$, on the transport properties of  graphene pn junctions. We use the non-equilibrium Green function (NEGF) approach with a nearest neighbor tight-binding description of graphene. This method allows us to accurately simulate the pn junction conductance for both abrupt and graded junctions under different bias conditions.   The value of $D_{w}$ is not known a priori because it depends on charge screening and the gate potential as governed by the Poisson equation. In this work, we employ the analytical screening model derived in \cite{zhang08}. Our numerical result based on the assumption of ballistic transport is in reasonable quantitative agreement with the experiments reported in \cite{huard07}.  As has been noted previously, we also found that the conductance follows an inverse square root dependence on $D_{w}$ when $D_{w}$ is large \cite{cheianov06}, but we find strong deviations from this trend occur when $D_{w}$ is sufficiently small.  Understanding this regime of operation has practical importance because devices based on the electron focusing property of  graphene pn junctions, i.e. graphene lenses \cite{cheianov07,shytov07} and filters \cite{katsnelson06}, are expected to operate in this regime.

\section{\label{sec:level2}GRAPHENE PN JUNCTIONS AND RESISTANCE ASYMMETRY}
\begin{figure}[t]
\centering
\scalebox{0.43}[0.43]{\includegraphics*[viewport=100 62 690 510]{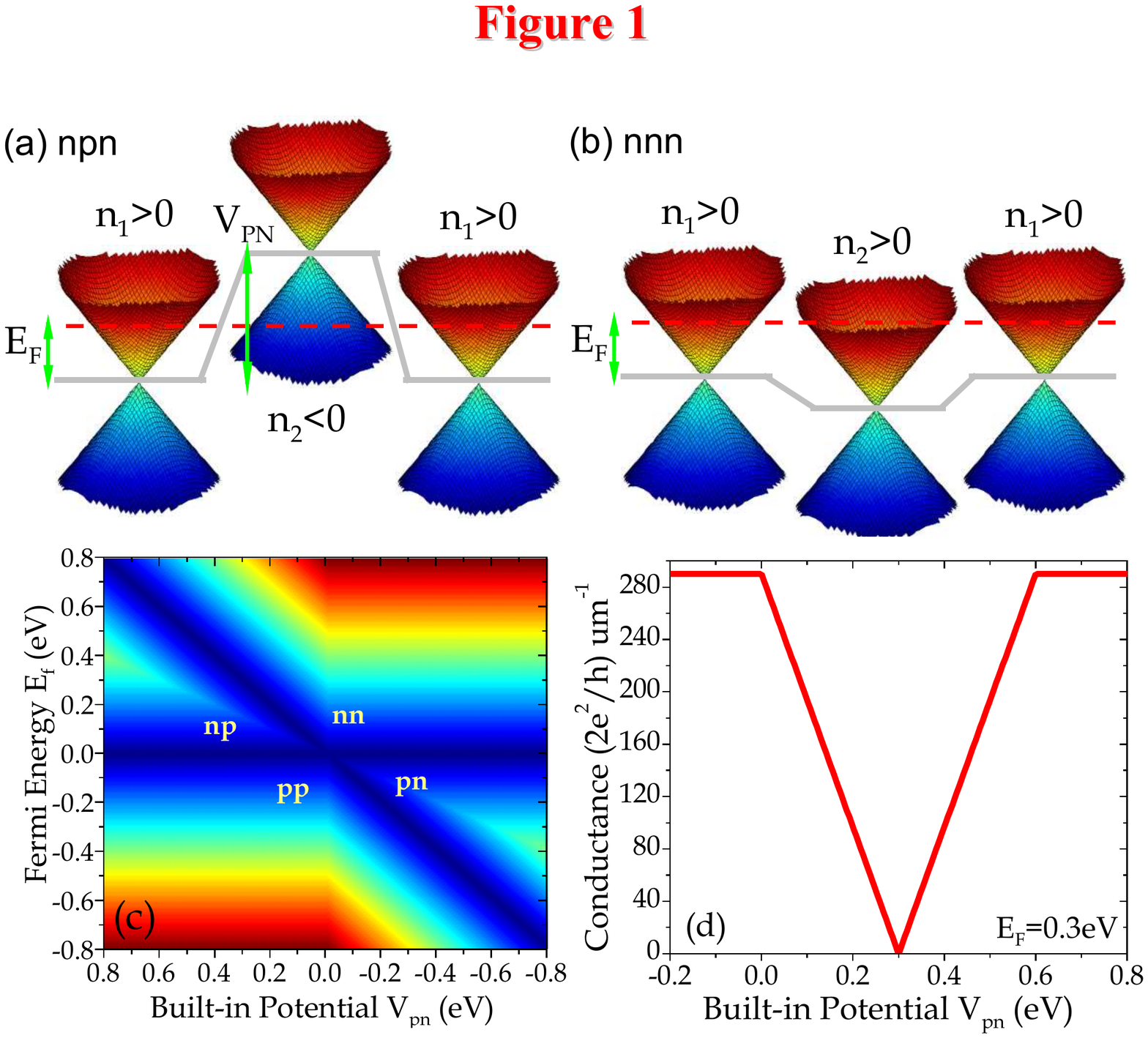}}
\caption{\footnotesize The energy band diagram of graphene npn and nnn devices depicted in (a) and (b) respectively. (c) is an intensity plot of the pn junction conductance, $\sigma_{pn}$, as a function of $E_{f}$ and $V_{pn}$ using Eq. \ref{simpledos2} and (d) plots $\sigma_{pn}$ as a function of $V_{pn}$ at $E_{f}$=$0.3eV$ using Eq. \ref{simpledos2}. }
\label{fig1}
\end{figure}

Before describing the simulation method used in this study, we define some terms that will be used in subsequent discussions, present a simple picture of the conductance of a graphene npn or pnp junction, and identify the issues that will be addressed in the numerical study. Figure \ref{fig1}(a,b) are depictions of the energy band diagram for the experiment, which shows the location of the intersection of the conduction and valence bands (the so-called Dirac point) vs. position.  A back gate controls the location of the Fermi level, $E_{f}$, in the source/drain regions.  For $E_{f}$$>$$0$ (above the Dirac point), there is an increase of electrons with respect to $E_{f}$=$0$ , the electron density is greater than zero, $n_{1}$$>$$0$, and the material is n-type.  If the back gate is biased negative so that $E_{f}$$<$$0$, then $n_{1}$$<$$0$ and the source/drain regions are p-type.  Similarly, the channel (middle) region can be biased by a top gate to be either p-type ($n_{2}$$<$$0$) or n-type ($n_{2}$$>$$0$). The top gate also controls the built-in potential, $V_{pn}$, of the np/nn junction. Thus, appropriate top and bottom gate voltages can produce npn, nnn, pnp, or ppp structures.  Near equilibrium conditions are assumed, so that the source and drain Fermi energies are the same. In this paper, we shall assume that the channel length is greater than the carrier's phase coherent length, allowing us to ignore wave interference effects within the channel. Therefore, we can reduce the problem by only studying the transport across a single pn junction.

As a first step to understanding the near-equilibrium conductance of a pn junction, $\sigma_{pn}$ ($S/um$), as a function of $V_{pn}$ and $E_{f}$ conditions, we adopt a simple density-of-states argument in the Landauer picture. In this simple analysis, $\sigma_{pn}$ can be written in the following form,
\small
\begin{eqnarray}
\sigma_{pn}=\frac{2e^2}{Wh}min\left(M_{1},M_{2}\right)
\label{simpledos2}
\end{eqnarray}
\normalsize
where $M_{1/2}$ is the number of modes in the source/channel respectively and $W$ is the device width. Eq. \ref{simpledos2} mimicks the matching of transverse momentum between the source and channel in a ballistic manner, such that the current will be limited by the region with the smaller number of modes. \textcolor{black}{In this simple exercise, we assume a zero temperature treatment, so that $M_{1/2}$ refers to the number of modes at their respective $E_{f}$. }

Figure \ref{fig1}(c) is an intensity plot of $\sigma_{pn}$ computed using Eq. \ref{simpledos2} versus $V_{pn}$ and $E_{f}$. The dark blue lines are regions of low current.  The horizontal line occurs when the source Fermi level is at the Dirac point. Since $n_{1}$=$0$, so no current flows.  The diagonal black line described by $V_{pn}$=$E_{f}$ occurs when the channel region is adjusted to place the Dirac point in the channel at the Fermi energy of the carrier. Since $n_{2}$=$0$  along this diagonal line, no current flows.  Very similar features are observed experimentally \cite{huard07,stander08}.  Looking more closely, we can plot  $\sigma_{pn}$  vs. $V_{pn}$ at a fixed $E_{f}$=$0.3eV$.  As shown in Fig. \ref{fig1} (d), the conductance vs. $V_{pn}$ is symmetrical about $V_{pn}$=$E_{f}$. In experiments, an asymmetry about $V_{pn}$=$E_{f}$ is observed \cite{huard07,stander08}. This asymmetry cannot be captured by simple density-of-states argument because its origin is quantum mechanical in nature i.e. quantum tunneling and wavefunction mismatch. Our goal in this paper is to explain the role that quantum tunneling and wavefunction mismatch played in the observed asymmetry and then quantitatively explain the magnitude of the asymmetry observed in experiments.

\section{\label{sec:level3}SIMULATION METHODS}

\begin{figure}[t]
\centering
\scalebox{0.39}[0.39]{\includegraphics*[viewport=100 101 700 490]{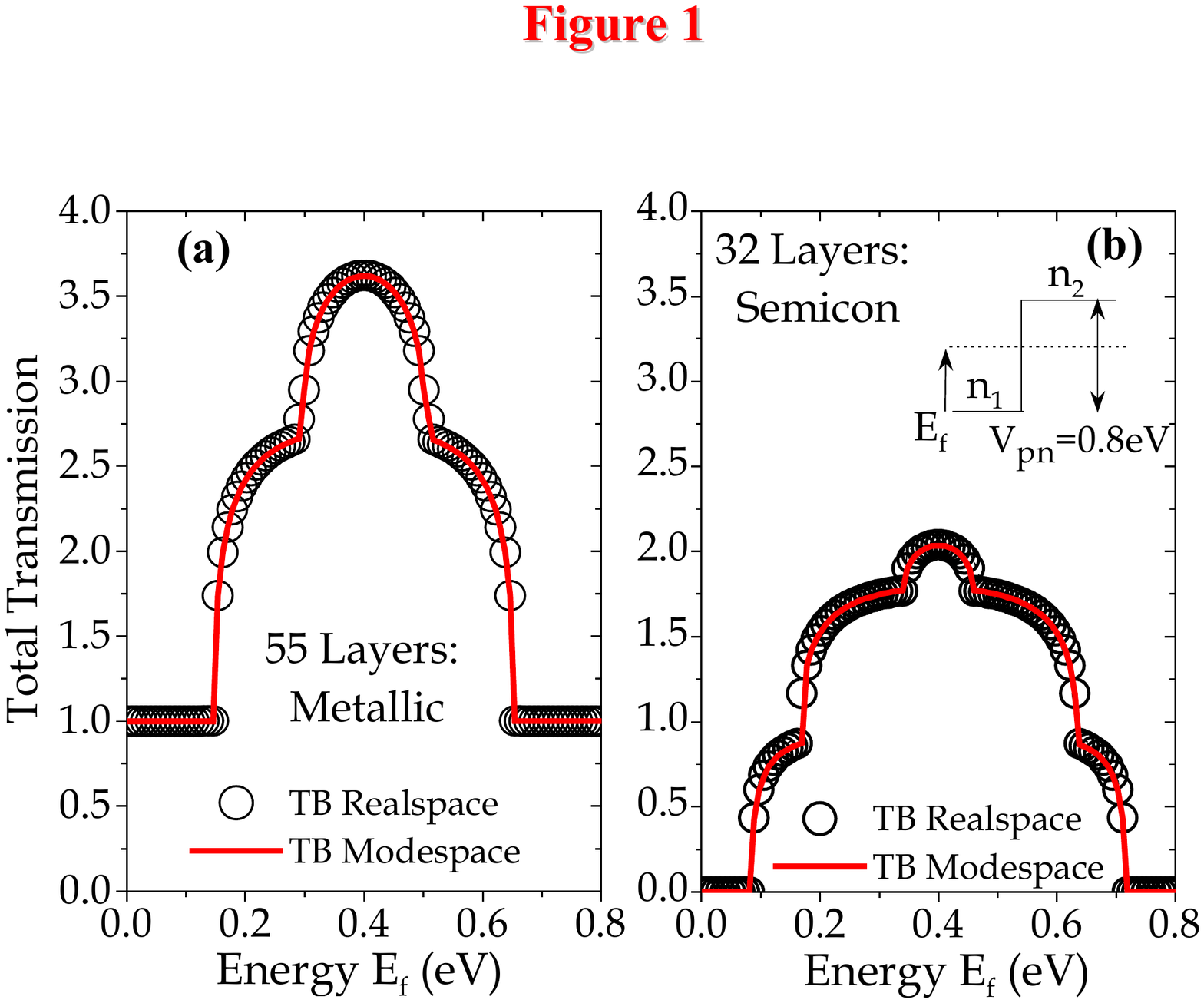}}
\caption{\footnotesize Total transmission as a function of energy calculated for (a) $55$ layers ($W\approx 13.7nm$) metallic armchair ribbons and (b) $32$ layers ($W\approx 8nm$) semiconducting armchair ribbons transporting across a pn junction with a built-in potential of $V_{pn}$=$0.8eV$. In both cases, we compared the method outlined in our paper (solid line) with that of real-space tight-binding approach (open symbols).}
\label{fig2}
\end{figure}

In this section, we briefly outline the NEGF formalism for quantum ballistic transport and describe its application to non-equilibrium transport \cite{datta97,haug96} through graphene pn junctions \footnote{Note that we are using the NEGF method in the ballistic limit where it is equivalent to the Landauer approach, as explained in chapter 8 of \cite{datta97}. }.  The central quantity in the theory is the retarded Green's function,
\small
\begin{eqnarray}
G(\epsilon,k_{y})=\left[(\epsilon+i\delta) I-H(k_{y})-U-\Sigma_{l}(\epsilon,k_{y})-\Sigma_{r}(\epsilon,k_{y})    \right]^{-1}
\end{eqnarray}
\normalsize
where $\delta$ is an infinitesimal quantity in the channel region but is adjusted to provide a non-vanishing density-of-states at the Dirac point for the contact regions \cite{henning07,mojarad07}. The Hamiltonian is formulated by treating only nearest-neighbor interaction between the $p_z$ orbitals \cite{wallace47,saito98}.   Assuming that the device width is large and homogeneous along the direction transverse to current flow, we can write the Hamiltonian as
\small
\begin{eqnarray}
H=\left[
\begin{array}{cccc}
\alpha & \beta_{1} & & \\
\beta_{1}^{\dagger} & \alpha & \beta_{2} & \\
& \beta_{2}^{\dagger} & \alpha & \ddots \\
&& \ddots & \ddots
\end{array}\right]
\end{eqnarray}
\normalsize
where $\alpha$, $\beta_{1}$ and $\beta_{2}$ are all $2\times 2$ matrices given by,
\small
\begin{eqnarray}
\begin{array}{ccc}
\alpha=\left[
\begin{array}{cc}
0 & t_{c} \\
t_{c} & 0
\end{array}\right]&
\beta_{1}=\left[
\begin{array}{cc}
0 & 0 \\
t_{y}^{\dagger} & 0
\end{array}\right]&
\beta_{2}=\left[
\begin{array}{cc}
0 & 0 \\
t_{y} & 0
\end{array}\right]
\end{array}
\end{eqnarray}
\normalsize
\normalsize
where $t_{c}$ is the nearest neighbor orbital coupling energy and \textcolor{black}{$t_{y}$=$t_{c}+t_{c}e^{ik_{y}a_{0}}$. The lattice parameter, $a_{0}$=$\sqrt{3}a_{cc}$ , where $a_{cc}$=$1.44\AA$ is the c-c bond distance. The quantum number $k_{y}$ is the quantized transverse momentum, to be elaborated upon in next paragraph.} The contacts' self-energies, $\Sigma_{i}(\epsilon,k_{y})$, are obtained by solving $\Sigma_{i}$=$\tau_{i} g_{i} \tau^{\dagger}_{i}$ where $g_{i}$ is the surface green function associated with the contact.  \textcolor{black}{We should mention that for armchair ribbon an analytical closed form solution for $g_{i}$ is possible, since the wavefunction is known, both in the tight-binding \cite{zheng07} and Dirac formalisms \cite{brey06}. However, in our numerical treatment, we had employed a non-negligible $\delta$ in the contact regions, which prevent us from using the analytical closed form solution for $g_{i}$. Therefore, $g_{i}$ is computed numerically with an iterative scheme proposed in \cite{sancho84}.} Finally, the current through contact $i$ can then be computed using,
\small
\begin{eqnarray}
I_{i}(\epsilon)=\frac{2q}{h}\sum_{k_{y}} trace\left[\Sigma^{in}_{i}(\epsilon)A(\epsilon)-\Gamma_{i}(\epsilon)G^{n}(\epsilon)\right]
\label{currenteq}
\end{eqnarray}
\normalsize
where $A$=$i(G-G^{\dagger})$ is the local density-of-states, $\Sigma^{in}_{i}$=$f_{0}(\epsilon)\Gamma_{i}(\epsilon)$ is the filling function (analogous to the in-scattering function for incoherent case), $f_{0}(\epsilon)$ is the Fermi function of the contacts, and $\Gamma_{i}$=$i(\Sigma_{i}-\Sigma_{i}^{\dagger})$ is the contact broadening factor.  In Eq. \ref{currenteq}, $G^{n}(\epsilon)$ is the electron correlation function given by $G(\Sigma^{in}_{l}+\Sigma^{in}_{r})G^{\dagger}$. $G$ and $G^{n}$ are computed using the recursive green function algorithm through Dyson's equation \cite{lake97}, and making use of the fact that the Hamiltonian is tridiagonal in nature.

By assuming an armchair ribbon configuration and imposing a box-boundary condition and solving the Dirac equation, Brey and Fertig showed \cite{brey06} that the transverse momentum is quantized according to
\small
\begin{eqnarray}
k_{y}=\left(\frac{2\pi}{3a_{0}}+\frac{2\pi n}{2W+a_{0}}\right)\pm \frac{2\pi}{3a_{0}}
\label{breyeq}
\end{eqnarray}
\normalsize
for all integer $n$ and $W$ is the width of the device. The last term accounts for the momentum of the Dirac points, $\vec{K}$ and $\vec{K}'$, where the upper/lower sign is used when $n$ is even/odd respectively. 

Treating the problem in terms of transverse modes greatly reduces the computation burden while still providing accurate results when the potential in the transverse direction is uniform \cite{venugopal02,guo03,zhao09}.  This approach essentially translates a two-dimensional real space transport problem into $m$ decoupled one-dimensional real space transport problem, where $m$ is the number of relevant transverse modes. In the limit of large device width, $W$, we have $m$$\propto$$W\epsilon$. \textcolor{black}{Fig. \ref{fig2} compares this mode space method with two-dimensional real space NEGF calculations for a 55-layer metallic graphene ribbon and a 32-layer semiconducting ribbon. `n-layer' refers to the number of layers of carbon atoms along the width direction, where $W=na_{0}$.} In both cases, the two methods give nearly identical energy-resolved transmission functions for transport across a pn junction.  The mode space approach has a computational burden that scales linearly with device width, which makes it possible to study typical experimental npn-type structures.

\begin{figure}[t]
\centering
\scalebox{0.44}[0.44]{\includegraphics*[viewport=109 50 735 540]{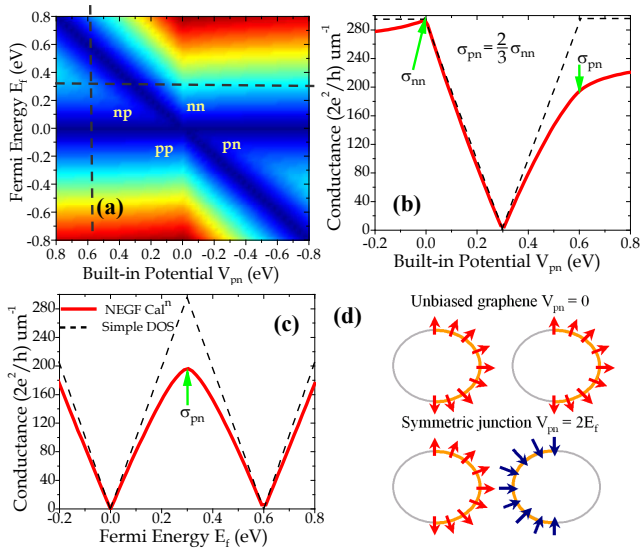}}
\caption{\footnotesize Simulation results from NEGF calculation of an abrupt graphene pn junction with device width of $0.5um$ at temperature of $4K$. (a) shows the intensity plot of conductance as function of built-in potential $V_{pn}$ and Fermi energy $E_{f}$ (see inset of figure \ref{fig2}(b) for definitions). Blue/red denotes low/high intensity respectively. (b) and (c) plots the conductance curves as function of $V_{pn}$ (at $E_{f}$=$0.3eV$) and $E_{f}$ (at $V_{pn}$=$0.6eV$) respectively. The dashed lines indicates the estimated conductance from a simple density-of-states argument (see text) for each case. In each case, the conductance of an unbiased graphene ($\sigma_{nn}$) and a symmetric pn junction ($\sigma_{pn}$) are indicated. (d) depicts the constant energy contour and its pseudo-spin alignment at each side of the junction for an unbiased graphene and a symmetric pn junction.  }
\label{fig3}
\end{figure}
\begin{figure*}[t]
\centering
\scalebox{0.6}[0.6]{\includegraphics*[viewport=35 245 745 488]{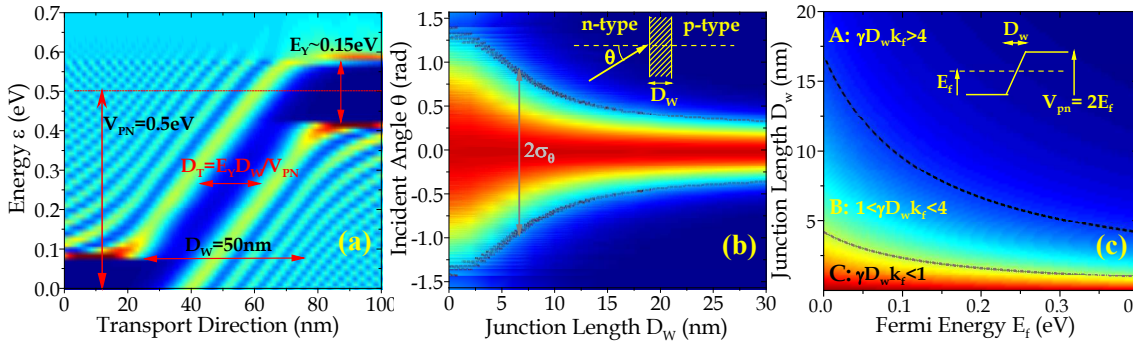}}
\caption{\footnotesize \bfseries Analysis of the impact of junction width on conductance modulation of a symmetric pn junction. \normalfont (a) plots the energy-resolved local density-of-states (i.e. $G\Sigma^{in}G^{\dagger}$) for a $W$=$0.5um$ pn junction device with a junction length of $D_{w}$=$50nm$ and a built-in potential $V_{pn}$=$0.5eV$.  This is generated for the transverse mode with a transverse energy of $\epsilon_{y}$$\approx$$0.075eV$, which yields an apparent `bandgap' of $\approx$$0.15eV$ as seen in the plot. $\epsilon=0$ is set at the Dirac point of the source (b) is the intensity plot of the transmission as a function of incident angle $\theta$ and $D_{w}$. The device is a symmetric pn junction with $V_{pn}$=$2E_{f}$, where $E_{f}$=$0.3eV$. An illustration of the setup is shown in the inset. (c) plots the fraction of conductance contribution due to wavefunction mismatch, i.e. $\sigma_{pn}^{wfm}/\sigma_{pn}$, for a symmetric pn junction (see Eq. \ref{spintunn}). }
\label{fig4}
\end{figure*}

\section{\label{sec:level4} RESULTS}
As discussed in Sec. II, the experimental setup \cite{huard07,williams07,ozyilmaz07,gorbachev08}  uses a back-gate and top-gate to control the source/drain contacts Fermi energy, $E_{f}$, and the potential difference across the pn junction, $V_{pn}$, respectively. The electron density in the source/drain regions is $n_{1}$ and in the channel, $n_{2}$; a negative value can be interpreted as a positive hole density. In this study, we assume that the applied source-drain bias is small, in accordance with experiments \cite{huard07}. 

We shall also assume that the channel length is greater than the carrier's phase coherent length, allowing us to ignore any quantum interference effects within the channel i.e. Fabry Perot effects. On this premise, the transport process across a pn junction on one side of the channel would not be influenced by the presence of the pn junction on the other side of the channel. Hence, we shall focus only on the transport across a single pn junction. We begin by examining the pn junction's conductance ($\sigma$ in units of $S/um$) when the junction is abrupt i.e. $D_{w}\approx 0$.

\subsection{\label{ss:level1}Conductance for an Abrupt PN Junction}
Fig. \ref{fig3}(a), which shows the computed conductance as function of $E_{f}$ and $V_{pn}$ for an abrupt pn junction, exhibits the features typically observed in experiments \cite{stander08,huard07}. The four distinctive regime of operations (i.e. nn, pp, np and pn) are partitioned by the $E_{f}$=$0$ and $E_{f}$=$V_{pn}$ lines, which correspond to conditions where either/both the contacts and channel are at zero equilibrium charge density as previously discussed in Se. II using simple density-of-states argument. Examining more closely, Fig. \ref{fig3}(b,c) plot the conductance as a function of $V_{pn}$ ($E_{f}$=$0.3eV$) and $E_{f}$ ($V_{pn}$=$0.6eV$) respectively. The red solid line is NEGF calculation while the dashed black line is calculated using Eq. \ref{simpledos2}. In Fig. \ref{fig3}(b), one can clearly see the conductance asymmetry with respect to the $V_{pn}$=$E_{f}$ point. This conductance asymmetry is an experimentally observed phenomena \cite{huard07} which cannot be captured by a simple density-of-states argument. 

Conductance asymmetry refers to the difference in conductance between a np junction and its nn counterpart. By `counterpart', we mean that $n_{2}$ for np junction is $-n_{2}$ for nn junction, while $n_{1}$ is the same for both junctions. In Fig. \ref{fig3}(b),  $\sigma_{nn}$ is the conductance of an unbiased graphene (i.e. $V_{pn}=0$), while $\sigma_{pn}$ is the conductance of a symmetric np junction (i.e. $V_{pn}=2E_{f}$). Clearly $\sigma_{nn}$$>$ $\sigma_{pn}$, a feature that is observed experimentally and captured by our NEGF calculation. The asymmetry is due to wave-function mismatch at the junction interface.

Electrons in graphene have a two-component wave function, which is often refered to as pseudo-spin in analogy to the two-component wave function for spins. \textcolor{black}{In the vicinity of the Dirac point, the two-dimensional Hamiltonian can be written in a form of Dirac equation \cite{ando98,semenoff84} i.e. $H= \upsilon_{f}\vec{\sigma}\cdot\vec{P}$, where $\vec{\sigma}$ and $\vec{P}$ are the Pauli spin matrices and momentum operator respectively. By convention, the definition of pseudo-spin is such that its direction is parallel to the group velocity, since the group velocity operator is defined as $\vec{\upsilon}_{G}=\nabla_{P} H= \upsilon_{f}\vec{\sigma}$. Fig. \ref{fig3}(d) provides an illustration of the constant energy contour for an unbiased nn graphene and symmetric pn junction case. The arrows simultaneously represents the group velocity and pseudospin, which points inwards/outwards for the valence/conduction band respectively. For the nn case, the velocity vectors are similarly aligned for each side of the junction. For the pn case, $\vec{\upsilon}_{G}$ changes sign across the junction. Analogous to spin, the wavefunction for the n and p regions can be expressed as $\left|\Psi_{n}\right\rangle=(1,e^{i\theta})/\sqrt{2}$ and $\left|\Psi_{p}\right\rangle=(1,e^{i(\theta+\pi)})/\sqrt{2}$ respectively (where $\theta=tan^{-1}(k_{y}/k_{f})$). The tranmision probability across the junction for a particular mode can then be written simply as $\left|\left\langle \Psi_{n}\right.\left|\Psi_{p}\right\rangle\right|^{2}$. For $k_{y}$=$0$, the wavefunction is perfectly matched i.e. $\left|\Psi_{n}\right\rangle=\left|\Psi_{p}\right\rangle$, so transmission is unity (i.e. Klein tunneling \cite{katsnelson06}). Through theoretical analysis, we derive $\sigma_{pn}=2/3\sigma_{nn}$, where the factor of $2/3$ is due to $\Sigma_{ky}(1-k_{y}^{2}/k_{f}^{2})$. Similarly, it can also be shown that the conductance when $V_{pn}\rightarrow\infty$ would approach the asymptotic value of $\approx$$(1/2+\pi/8)\sigma_{nn}$. Rigorous NEGF calculations  faithfully reproduces these features as shown in Fig. \ref{fig3}(b).}

In summary, the conductance of an abrupt pn junction can be understood as being controlled by the region in which the number of conducting channels is smallest. Wavefunction mismatch reduces the current by a factor of $2/3$ in a symmetric np junction compared to an unbiased graphene. These features are captured by NEGF simulation, but realistic junctions have a finite transition width, $D_{w}$, which, as we emphasize in the following section, plays an important fole. 

\subsection{\label{ss:level1}Effect of Junction Width on Symmetric PN Junction Conductance}

For realistic np junctions, the transition length across which the charge density changes monotonically from n-type to p-type is finite. The width of the junction transition region has a strong influence on the conductance of the junction. To understand these effects, it is instructive to consider each of the transverse modes individually. According to Eq. \ref{breyeq}, for a wide graphene sheet, the transverse momentum with respect to the Dirac point is given by $k_{y}\approx n\pi/W$ where $n$ is an integer. We can view each mode as a ray with an angle of incidence on the junction $\theta=tan^{-1}[k_{y}(k_{f}^{2}-k_{y}^{2})^{-0.5}]$. Fig. \ref{fig4}(a) plots the energy-resolved local density-of-states ($G\Sigma^{in}G^{\dagger}$) for a pn junction with a transition length of $D_{w}$=$50nm$ and a built-in potential $V_{pn}$=$0.5eV$. Only one transverse mode with a transverse energy of $\hbar v_{f}k_{y}$$\approx$$0.075eV$ is considered. The dark blue region correspond to a low density of states. Because of the quantization of transverse wavevectors, an apparent bandgap of $0.15eV$ is observed. Therefore, electrons at Fermi energy of $E_{f}$$<$$V_{pn}$ will have to quantum mechanically tunnel through this $k_{y}$-dependent bandgap when moving from one side of the junction to the other. \textcolor{black}{This problem is analogous to the classic band-to-band tunneling problem in direct band-gap semiconductors \cite{kane61}.}

Based on this physical picture, Cheianov and Falko \cite{cheianov06} worked out the WKB tunneling probability for a given transverse mode to be $e^{-\pi k_{f}D_{w}sin^{2}(\theta)/2}$. This tunneling expression is derived by assuming a symmetric pn junction. \textcolor{black}{In similar spirit to classic band-to-band tunneling treatment, the electric field across the pn junction is assumed to be linear, given by $2E_{f}/D_{w}$ in this work. Realistically, the potential energy landscape at the beginning/end of the pn junction would exhibits a more quadratic profile. However, only the details of the potential energy landscape within the tunneling region contributes to the WKB tunneling probability. In this linear electric field approximation, the tunneling distance is simply given by, $D_{t}=2\hbar v_{f}k_{y}D_{w}/V_{pn}$.} Fig. \ref{fig4}(b) shows the NEGF-computed electron transmission as a function of incident angle, $\theta$, and transition length, $D_{w}$, for a symmetric pn junction with $E_{f}$=$0.3eV$. As expected, increasing $D_{w}$ results in a decreased angular bandwidth ($2\sigma_{\theta}$) of the allowable transverse modes that can be transmitted across the pn junction, which subsequently leads to a decreased pn junction conductance. Based on the WKB tunneling forumla, it can be shown that $\sigma_{pn}$$\propto$$ \sqrt{k_{f}/D_{w}}$. This leads to the question of whether the junction rectification metric can be improved by using a larger $D_{w}$.  Unfortunately, $\sigma_{nn}$/$\sigma_{pn}$ which depends on $D_{w}$ in an inverse square root manner, yields only a factor of $10$ with $D_{w}\approx100nm$ at a typical $E_{f}$ of $0.3eV$. 

Device concepts based on the electron focusing property of pn junction, i.e. graphene lenses \cite{cheianov07,shytov07} and filters \cite{katsnelson06} operate best in a symmetric pn junction. These devices operate in a regime where $\sigma_{\theta}$ has to be as large as possible so as to reconstruct back a point source image on the other side of the pn junction. This implies that $D_{w}$ has to be sufficiently small to enhance tunneling at high incident angle. By accounting for both the wavefunction mismatch and tunneling factor in the following manner, $T(\theta)=cos^{2}(\theta)e^{-\pi k_{f}D_{w}sin^{2}(\theta)/2}$, we found that we are able reproduce the result of NEGF for arbitrary $D_{w}$ (not shown). By integrating $T(\theta)$ over all transverse modes, we can arrive at a more general result for the conductance of a symmetric pn junction (in units of $2\frac{e^{2}}{h}$),
\small
\begin{eqnarray}
\nonumber
\sigma_{pn}=\frac{\sqrt{k_{f}}}{\sqrt{\pi\gamma D_{w}}}erf(\sqrt{\gamma D_{w}k_{f}})+\\
\frac{2}{\pi^{2}D_{w}}e^{-\gamma D_{w}k_{f}}-\frac{\sqrt{2}}{\pi^{2}\sqrt{k_{f}}D_{w}^{1.5}}erf(\sqrt{\gamma D_{w}k_{f}})
\label{spintunn}
\end{eqnarray}
\normalsize
where $\gamma=\pi/2$. The first term in Eq. \ref{spintunn} is due to the tunneling factor. The last two terms are corrections due to wavefunction mismatch. Eq. \ref{spintunn} can be written as $\sigma_{pn}=\sigma_{pn}^{tun}+\sigma_{pn}^{wfm}$. Fig. \ref{fig4}(c) is an intensity plot of $\sigma_{pn}^{wfm}/\sigma_{pn}$ as a function of $E_{f}$ and $D_{w}$. The blue region represents $\sigma_{pn}^{tun}\gg\sigma_{pn}^{wfm}$ while the red regions indicate $\sigma_{pn}^{tun}\ll\sigma_{pn}^{wfm}$. Evidently, conductance modulation is predominantly due to wavefunction mismatch only when $\gamma D_{w}k_{f}<1$, which suggests that for the tunneling component not to limit electron focusing applications, a $D_{w}<5nm$ is required. 

In summary, the conductance asymmetry of a graphene pn junction is due to two quantum mechanical processes, wavefunction mismatch and the need to tunnel through an apparent bandgap induced by the quantization of transverse momentum. Increasing $D_{w}$ results in a decreased angular bandwidth ($2\sigma_{\theta}$) of the allowable transverse modes that can be transmitted across the pn junction. This leads to a decreased pn junction conductance and would eventually result in a larger magnitude of odd resistance $R_{odd}$.

\subsection{\label{ss:level3}Odd Resistance and Comparison With Experiments}

\begin{figure}[t]
\centering
\scalebox{0.34}[0.34]{\includegraphics*[viewport=50 135 745 480]{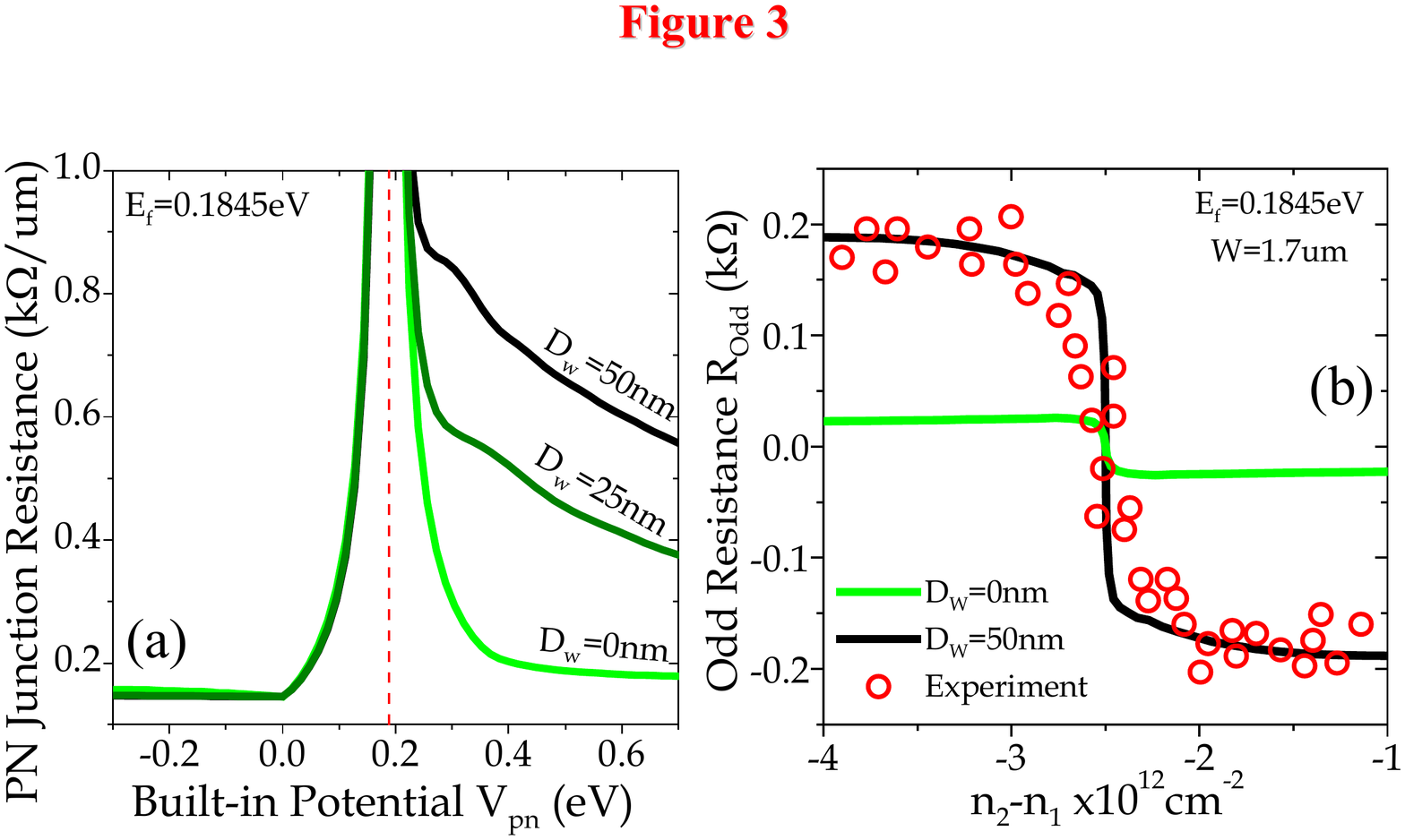}}
\caption{\footnotesize Theoretical resistance per unit width of a single pn junction as a function of $V_{pn}$ at $E_{f}$=$0.1845eV$ under different $D_{w}$ conditions (b) Comparison of NEGF and experimental odd resistance $R_{odd}$ for different $D_{w}$ plotting with respect to $n_{2}-n_{1}$. Experimental device has $W$=$1.7um$ with contacts Fermi energy also at $E_{f}$=$0.1845eV$.   }
\label{fig5}
\end{figure}

Finally, we shall examine the odd resistance, $R_{odd}$, of pn junction devices and compare our NEGF result with the experimental data reported in \cite{huard07}. Typically, the resistance asymmetry is characterized by analyzing the resistance of the device as a function of $V_{pn}$ at a given $E_{f}$. A quantity known as odd resistance, $R_{odd}$, can be obtained by taking the difference between the resistance of the npn and its nnn $\bold{counterpart}$ device i.e. $R_{odd}=\frac{1}{2}\left[R_{npn}-R_{nnn}\right]$, where the channel hole density for npn is equal to the electron density for nnn. Fig. \ref{fig1}(a,b) depicts the energy band of a typical npn device and its nnn $\bold{counterpart}$. $R_{odd}$ for a long channel device depends only on the odd resistance contribution from individual pn junctions. Essentially, $R_{odd}$ simply reduces to $R_{odd}=\left[1/\sigma_{np}-1/\sigma_{nn}\right]$ obtained by a simple sum of the resistance of the two adjacent pn junctions. As a first step, we shall investigate the contributions of wavefunction mismatch and quantum mechanical tunneling processes to the magnitude of $R_{odd}$. Fig. \ref{fig5}(a) shows the theoretical pn junction resistance as a function of $V_{pn}$ at $E_{f}$=$0.1845eV$ under different $D_{w}$ conditions. The odd resistance contribution due to wavefunction mismatch alone (i.e. $D_{w}$=$0$) does not adequately account for the $R_{odd}$ observed in experiments as evident in Fig. \ref{fig5}(b). In fact, it only accounts for $10\%$ of the $R_{odd}$. Accounting for finite $D_{w}$ is essential to match the experimental data.

\begin{figure}[t]
\centering
\scalebox{0.45}[0.45]{\includegraphics*[viewport=81 150 660 500]{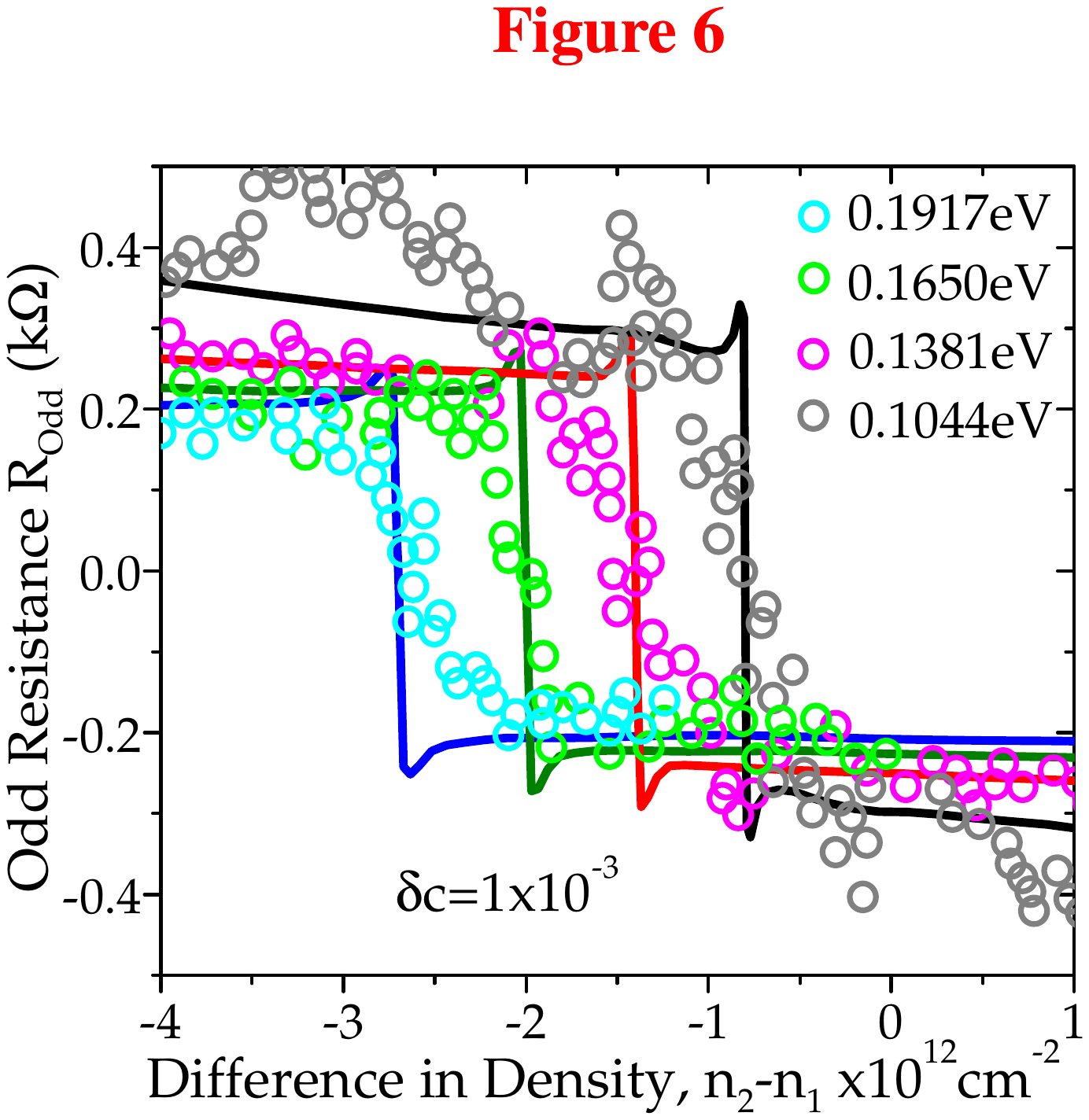}}
\caption{\footnotesize Comparison of the NEGF simulated $R_{Odd}$ with that of experimental results in \cite{huard07} at different Fermi energies i.e. $E_{f}$=$0.1917$, $0.165$, $0.1381$ and $0.1044eV$. The device width is $1.7um$ and operating temperature is $4K$. In the simulations, the contacts are assumed to have an energy broadening of $1meV$. The calculations assumed $\alpha$=$0.78$ due to a $\kappa$ of $~4.5$ \cite{huard07} and an oxide thickness of $T_{ox}$$\approx$$80nm$ as a fitting parameter.  }
\label{fig6}
\end{figure}

In our NEGF simulation, $D_{w}$ is the parameter that we need to determine prior to our NEGF calculations. The determination of $D_{w}$ is an electrostatics problem which is sensitive to the specific device geometry. In this work, we used the model presented in \cite{zhang08}, which expresses $D_{w}$ in terms of $n_{1}$ and $n_{2}$ as follows,
\small
\begin{eqnarray}
D_{w}\approx 0.196\times \frac{V_{pn}}{\hbar v_{f} \alpha^{1/3}}\left(1-\frac{n_{2}}{n_{1}}\right)^{\frac{4}{3}}\left|\frac{T_{ox}}{n_{2}}\right|^{\frac{2}{3}}
\label{dwmodel}
\end{eqnarray}
\normalsize
where $\alpha$=$e^{2}/(\kappa\hbar v_{f})$. $\kappa$ and $T_{ox}$ are the effective dielectric constant and thickness of the oxide between the top gate and graphene device respectively\footnote{\textcolor{black}{We shall also briefly discuss the validity of the analytical non-linear screening model of Eq. \ref{dwmodel}. As discussed in \cite{zhang08}, the accuracy of the analytical model is dependent on the physical parameter $\alpha$. The model would yield a possible deviation from  numerical results of $\approx 25\%$ when $\alpha\approx 0.9$ at the charge neutrality point within the transition region. We emphasize that the error is of an oscillatory nature with the oscillation amplitude deterioting when it moves away from the charge neutrality point. This would serve to alleviate the average error. While on the other limit, $\alpha\approx 0.1$ yields an excellent agreement with the numerical result. In the set of experiments we studied in this manuscript, $\alpha \approx 0.78$ due to a $\kappa\approx 4.5$. Therefore, the average error is intermediate of these two limits. However, when $\alpha$ is large (approaching 1), one would need to also account for electron exchange and correlation effects (see e.g. \cite{adam08}), making the problem numerically non-tractable.     } }. 

Fig. \ref{fig6} shows the computed $R_{odd}$ at different $E_{f}$ conditions as a function of $n_{1}-n_{2}$ and its comparison with experimental data. The NEGF result achieves quantitative agreement with the experimentally observed odd resistance. In particular, the increase in odd resistance with decreasing $E_{f}$, a puzzling feature in the experiments \cite{huard07}, is captured by the simulations. This occurs because of the increasing $D_{w}$ with decreasing $E_{f}$ i.e. smaller $n_{1}$ which results in an increase of pn junction screening length. We note conductance oscillations at the smallest $E_{f}$. These oscillations are likely due to interference effects within the device channel, which are more pronounce at $E_{f}$=$0.1eV$ due to the large $D_{w}$ which leads to an effectively shorter channel length. The `spikes' observed in the NEGF simulations when $R_{odd}$ crosses zero are due to the zero density-of-states at Dirac point. Such spikes are not observed in the experiments, probably due to the presence of spatial fluctuations (electron-hole puddles) when $E_{f}$ approaches the Dirac point \cite{hwang07}. By construction, our NEGF model does not account for these electron/hole puddles.

\section{\label{sec:level3}CONCLUSIONS}

In this paper we presented a numerical study of electron transport in graphene pn junctions.  We first presented a very simple minimum density-of-states (or conducting channels) model to account for the overall shape of the conductance vs. source carrier density and junction potential. Such a simple model does not capture the resistance asymmetry observed experimentally. We then use NEGF simulation to explore in detail the role of wave-function mismatch (also called pseudo-spin) and quantum mechanical tunneling through the junction transition region.  In particular, we examined deviations from the inverse square root dependence of $\sigma_{pn}$ on $D_w$ due to wave-function mismatch at small $D_w$.  Finally, we compared the simulations to a recent experiment and showed that the numerical model is in reasonable agreement with experiments and explain the increase in odd resistance with decreasing carrier density in the source. The novel features of graphene's electronic structure lead to interesting possibilities for new devices, and this study shows that NEGF simulation should provide a useful tool to explore and assess device concepts.

$\bold{Acknowledgement}$ We gratefully acknowledge support of the Nanoelectronic Research Initiative and the Network for Computational Nanotechnology. We also acknowledge Prof Jing Guo for useful discussions on NEGF simulation by mode-space methods \cite{zhao09} and Srikant Srinivasan for useful discussions and comments on this work. 



\end{document}